\documentclass[%
 aip,
 amsmath,amssymb,
 reprint,%
]{revtex4-1}

\usepackage{graphicx}
\usepackage{dcolumn}
\usepackage{bm}

\usepackage[utf8]{inputenc}
\usepackage[T1]{fontenc}
\usepackage{mathptmx}
\usepackage{etoolbox}
\usepackage{enumitem}
\usepackage{color}

\makeatletter
\def\@email#1#2{%
 \endgroup
 \patchcmd{\titleblock@produce}
  {\frontmatter@RRAPformat}
  {\frontmatter@RRAPformat{\produce@RRAP{*#1\href{mailto:#2}{#2}}}\frontmatter@RRAPformat}
  {}{}
}%
\makeatother
\begin{document}

\preprint{AIP/123-QED}

\title{Experimental Sensitivity Enhancement of a Quantum Rydberg Atom-Based RF Receiver with a Metamaterial GRIN Lens}

\author{Anton Tishchenko}
\author{Demos Serghiou}
\author{Ashwin Thelappilly Joy}
\affiliation{5G $\&$ 6G Innovation Centre (5GIC $\&$ 6GIC), Institute for Communication Systems (ICS), University of Surrey, Guildford, GU2 7XH, U.K.}
\email{a.tishchenko@surrey.ac.uk, g.gradoni@surrey.ac.uk}
\author{Paul Marsh}
\author{Paul Martin}
\affiliation{PA Consulting Global Innovation and Technology Centre, Back Ln, Melbourn, Royston, SG8 6DP,~U.K.}
\author{Tim Brown}
\author{Gabriele Gradoni}
\author{Mohsen Khalily}
\author{Rahim Tafazolli}
\affiliation{5G $\&$ 6G Innovation Centre (5GIC $\&$ 6GIC), Institute for Communication Systems (ICS), University of Surrey, Guildford, GU2 7XH, U.K.}





\date{\today}

\begin{abstract}

\section*{\label{sec:level1}Abstract}
We experimentally demonstrate enhanced sensitivity of an atom-based Rydberg radio frequency (RF) receiver integrated with a gradient refractive index (GRIN) Luneburg-type metamaterial lens. By analyzing the electromagnetically induced transparency (EIT) effect in Cesium vapor, we compare receiver performance with and without the GRIN lens under a 2.2~GHz and a 3.6~GHz far-field excitation. Our measurements reveal a significant amplification of the EIT window when the lens is introduced, consistent with the theoretical prediction that the local E-field enhancement at the vapor cell reduces the minimum detectable electric field and improves the microwave electric field measurement sensitivity of the Rydberg atom-based RF receiver over an ultrawide bandwidth of the lens. This experimental validation demonstrates the potential of metamaterial-enhanced quantum RF sensing for a wide range of applications, such as electromagnetic compatibility (EMC) testing, quantum radar, and wireless communications.
\end{abstract}

\maketitle







Atom-based radio frequency (RF) sensing utilizes the effect of electromagnetically induced transparency (EIT) in a non-linear quantum optical effect \cite{bHarris}. EIT is used to excite the outermost electron, typically in a group 1 Alkali metal (such as Rubidium, Rb, or Cesium, Cs, which was used in this work), into a Rydberg state via a two-photon optical transition. Once the atom is excited to a Rydberg state, its unique optical and atomic properties enable the analysis of various phenomena, including studies of atomic structure and highly sensitive detection of electromagnetic waves \cite{bCai2022, bHolloway2014}, terahertz imaging \cite{bKevin2016}, quantum memory \cite{bKuzmich}, slowed light \cite{bSlowlight}, quantum computing \cite{bGiraldo}, electric field metrology \cite{bEMC}, quantum radar \cite{bRadar}, and wireless communication \cite{bMenchetti, bOtto2021}. \par
While the theory of atom-based RF sensing promises extremely wide bandwidths, Planck constant level of precision, and a sub-noise floor signal detection, its real-world performance is hindered by the Doppler effect between multi-body Rydberg atoms, that is always present in hot atomic vapor cells (i.e., room-temperature practical systems), resulting in the receiver’s sensitivity being far from its theoretical expectation \cite{bSensitivity}. Several methods have been considered in the literature to improve the sensitivity of bare Rydberg atom-based RF receivers, such as the use of a split ring resonator (SRR) \cite{bSRR} to capture the incident RF field and enhance it at the resonant frequency of the SRR, or other types of resonant structures \cite{bSRR2, bSRR3}. Despite being very effective, these methods cannot be considered suitable for electric field metrology due to inevitable spurious and harmonic emissions caused by resonating metals. Additionally, the narrowband nature of the resonating frequency further limits its potential applications. Other techniques consider the application of a Mach-Zehnder interferometer \cite{bLasers1} and combined laser arrays with cascaded diffraction gratings \cite{bLasers2}. However, such techniques are complex and expensive to implement, while their gains are also confined to a narrow RF bandwidth. \par
Therefore, we propose a fundamentally different approach for enhancing the sensitivity of a Rydberg RF receiver in this work, which has not been explored in the literature to the best of our knowledge. We propose encapsulating its vapor cell with a passive gradient refractive index (GRIN) metamaterial lens that does not consume active power nor create any spurious emissions. Since the receiver is expected to be in the far-field of the RF emission or wireless communication signal source, we consider a Luneburg-type GRIN lens that focuses planar wavefronts on a focal point, assumed to be at the center of the vapor cell, with a high gain and an ultrawide RF bandwidth. In this design, graded index profile arises via engineered subwavelength structure rather than continuous homogeneous optics. We designed, fabricated, and characterized the lens in an anechoic chamber. Then, we performed experimental measurement of the EIT window with a Rydberg RF receiver, reporting a significant improvement in its sensitivity. 

\begin{figure*}
\includegraphics[width=\textwidth]{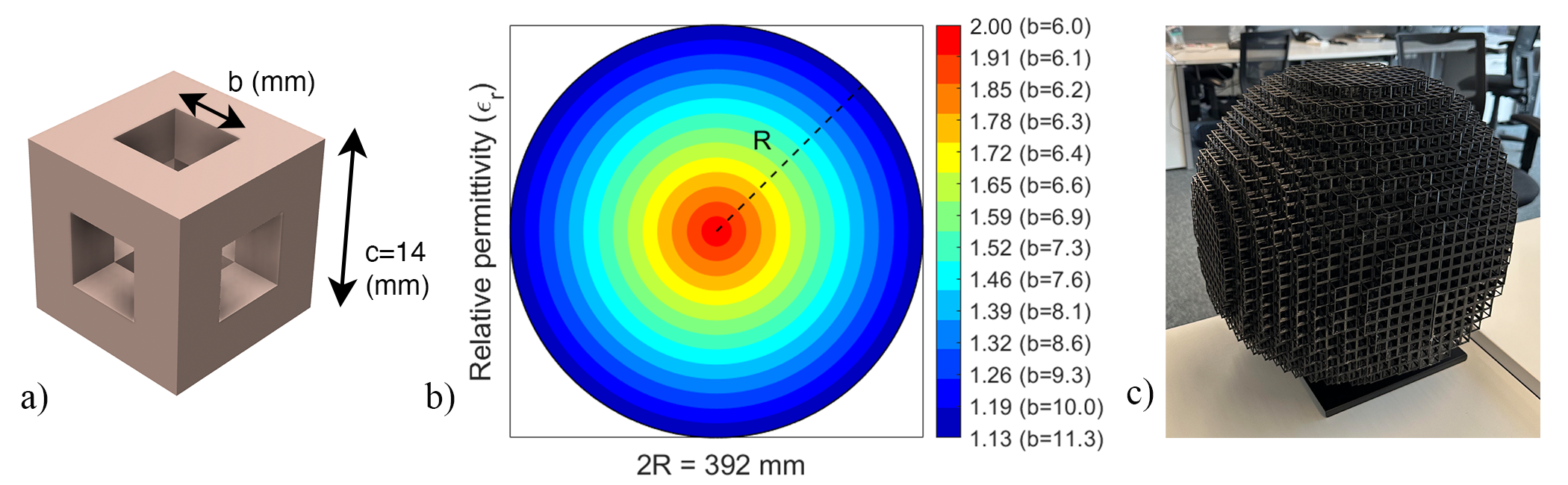}
\caption{\label{fig:lens}GRIN Luneburg-type metamaterial design, centred at 3.5~GHz, showing: (a) the unit cell with size $c=({\lambda}/6)^3\approx(14\text{mm})^3$ and the geometrical variable $b$,  (b) the refractive index $n$ variation as a function of $b$, simulated in CST Microwave Studio, and c) the assembled GRIN Luneburg-type metamaterial lens, consisting of eight 3D-printed fragments and made from the PLA material.}
\end{figure*}

\begin{figure*}
\includegraphics[width=0.9\textwidth]{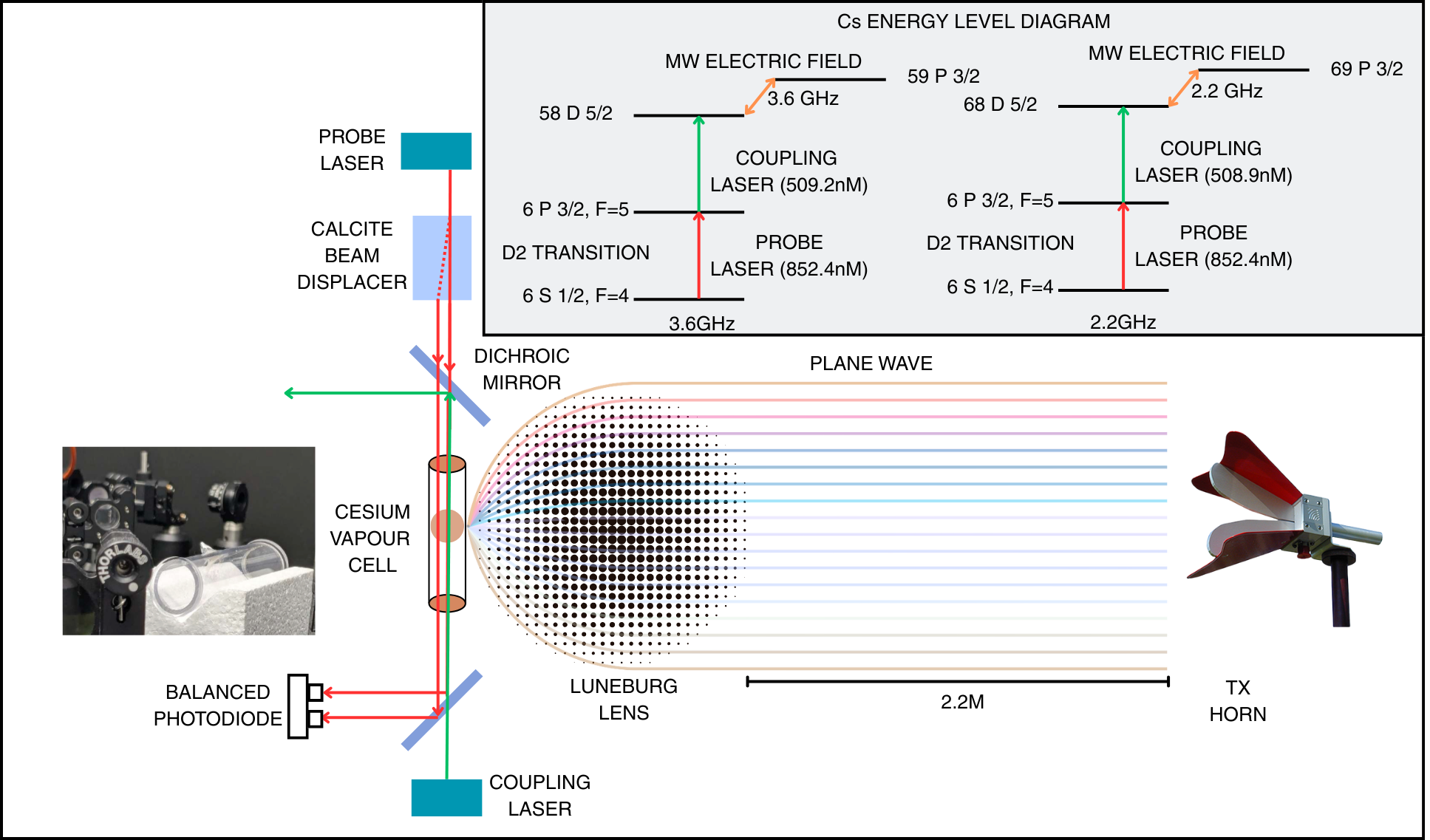}
\caption{\label{fig:setup}Schematic diagram of the proposed experiment, showing the Rydberg receiver with a metamaterial GRIN lens, including Cesium energy level diagrams at 2.2~GHz and 3.6~GHz.}
\end{figure*}

\begin{figure*}
\includegraphics[width=\textwidth]{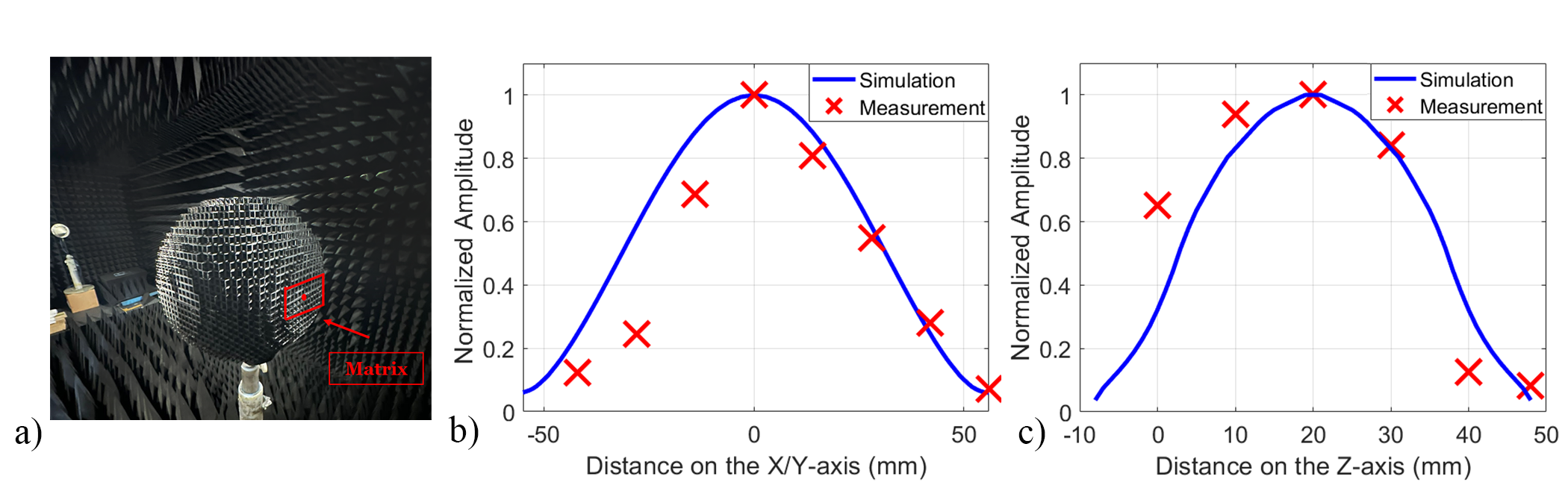}
\caption{\label{fig:chamber}Anechoic chamber testing, showing: a) test setup, b) x-axis (beam waist) measurement vs. CST simulation at 3.6~GHz, and c) z-axis (focal length) measurement vs. CST simulation at 3.6~GHz.}
\end{figure*}

\begin{figure*}
\includegraphics[width=\textwidth]{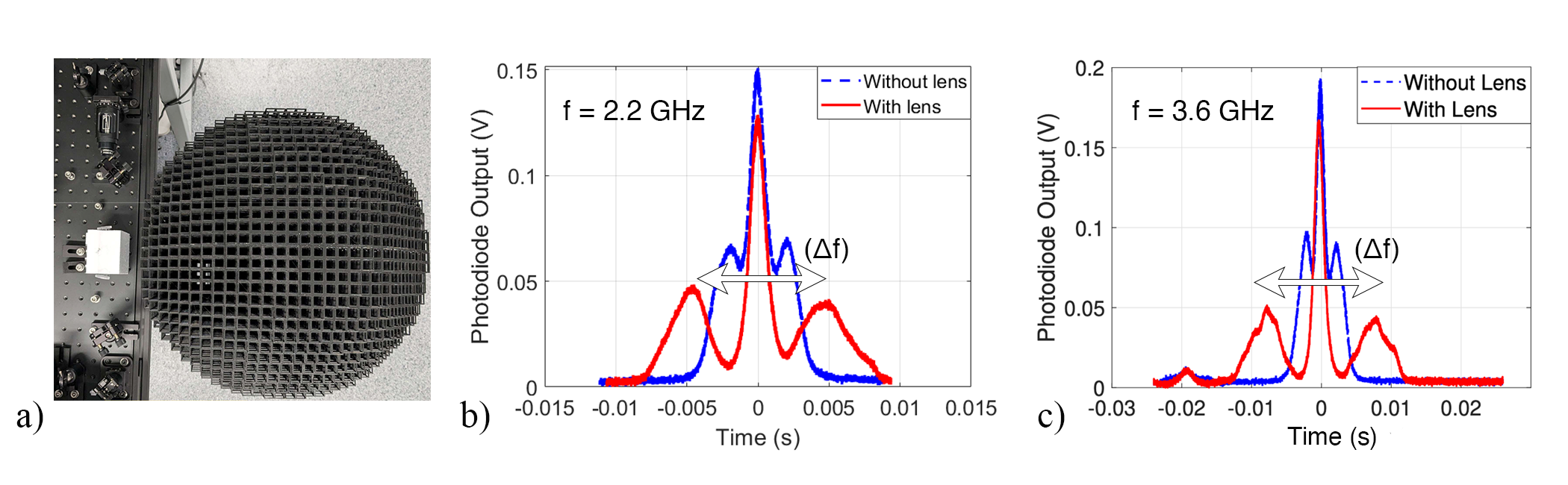}
\caption{\label{fig:eit}Experimental results, showing: 
(a) the Rydberg receiver and a GRIN Luneburg-type metamaterial lens test setup, b) the EIT window enhancement with and without the lens at 2.2~GHz, and (c) the EIT window enhancement with and without the lens at 3.6~GHz.}
\end{figure*}

For a two-state Rydberg RF receiver, the minimal detectable electric field corresponding to the quantum shot noise in the Autler–Townes limit \cite{bBussey,bShot2015} can be expressed as:
\begin{eqnarray}
E_{\text{min}}^{\text{no~lens}} & = & \frac{h}{\mu_{\text{RF-FI}}~T_m~\sqrt{N_m}}\;, 
\label{eq:e-field}
\end{eqnarray}
where $h$ is Planck's constant, $\mu_{\text{RF-FI}}$ is the RF transition dipole moment expressed in Coulomb-meter, $T_m$ is the measurement time, and $N_m$ is the number of independent measurements.  \par
We can see from (\ref{eq:e-field}) that the sensitivity of measurement is largely influenced by $\mu_{\text{RF-FI}}$, where a larger spacing between EIT peaks (i.e., split) would increase the sensitivity of the Rydberg receiver. Note that in the described experiment, the size of the vapor cell was considerably smaller than the wavelength of the carrier signal, hence we assume the effects of vapor cell electrometry on $\mu_{\text{RF-FI}}$ to be negligible.\par
In this work, we consider a Luneburg-type gradient index (GRIN) lens with refractive index variation $n(r) = \sqrt{\left(2-\left(r/R\right)^2\right)}$ in a general radial coordinate $(r)$ that has the electric field focusing ability expressed analytically per \cite{bMaci}:
\begin{eqnarray}
\left|\mathbf{E}(\rho)\right| & = & \frac{k\eta_0\left|\mathbf{\hat{s}_A} \times 
\left(\mathbf{\hat{s}_A} \times \mathbf{J_0}\right)\right| }
{4\pi\sqrt{R}\left(R^2-\rho^2\right)^{1/4}}\;, 
\label{eq:luneburg}
\end{eqnarray}
where $\rho$ is the radial offset from the center of the lens, approaching the focal region on the surface $\left(\lim_{\rho \to R^{-}}\right)$, and in this context designed to be in the centre of the vapor cell, $R$ is the radius of the lens, $k$ is the free-space wavenumber $\left(k=2\pi/\lambda\right)$, $\eta_0$ is the intrinsic impedance of free space, $\mathbf{J_0}$ is the equivalent current distribution on an aperture (per Huygens’ principle), and $\mathbf{\hat{s}} =\nabla \psi/n$ with the eikonal function $\psi(\mathbf{r})$ corresponding to the optical length. \par
Applying (\ref{eq:luneburg}) to the Rydberg receiver design implies that $\left|\mathbf{E}(\rho)\right|$ should be maximized by the physical properties of the GRIN lens to increase the local field amplitude at the vapor cell by a focusing factor $\gamma$ expressed as a linear focusing gain:
\begin{eqnarray}
\gamma & = & \frac{\left|\mathbf{E}(\rho)\right|}
{\left|\mathbf{E}_{\text{inc}}(\rho)\right|}\;, 
\end{eqnarray}
where $\mathbf{E_{inc}}(\rho)$ is the incident electric field at the center of the vapor cell without the GRIN lens. \par 
Then, assuming aligned polarization, we can express the enhancement of the Autler–Townes splitting $\left(\triangle f\right)$ in Hz by the GRIN lens with a linear focusing gain $\gamma$ as:
\begin{eqnarray}
\Delta f & = & \gamma\,\frac{\mu_{\text{RF-FI}}\,\left|\mathbf{E}_{\text{inc}}(\rho)\right|}{h}\;. 
\end{eqnarray}
Consequently, we can analytically observe that the minimum detectable external field in (\ref{eq:e-field}) improves by a factor of $1/\gamma$:
\begin{eqnarray}
E_{\text{min}}^{\text{with~lens}} & = & \frac{h}{\gamma~ \mu_{\text{RF-FI}}~T_m~\sqrt{N_m}}\;. 
\label{eq:enchanced-field}
\end{eqnarray}

Based on Snell’s law, when an electromagnetic wave passes between two media with different refractive indices, the transmitted wave bends towards or away from the normal. A metamaterial GRIN Luneburg-type lens exploits this by using a varying refractive index, typically from 1 at the edge to 2 at the center, to transform an incoming plane wave into circular wavefronts. Since natural GRIN materials do not exist, the lens is realized by discretizing the refractive index radially and assigning each unit cell a value from (\ref{eq:luneburg}). With a fixed unit-cell volume, the effective index is set by the dielectric fill fraction (parameter $b$). To avoid spurious emissions, polyactic acid (PLA) with refractive index $\approx2.99$ at 3.5 GHz was chosen. The simulation result in Fig. \ref{fig:lens} shows that indices between 1 and 2 are achieved by tuning the parameter $b$, with $\rho$ set to 216~mm. To simplify the fabrication, the spherical lens was divided into cubical lattices of volume $c=({\lambda}/6)^3\approx(14~\text{mm})^3$, where ${\lambda=84~\text{mm}}$ at 3.5 GHz. The overall lens diameter, $2R = 392$ mm, was chosen to be $\approx4.5 \times\lambda$. It was fabricated in eight segments on Bambu Lab X-1 Carbon printers (100\% infill, 0.2 mm layer thickness) and assembled into a sphere. 

When an RF wave passes through the GRIN lens, its phase is adjusted to focus at the vapor cell center, producing a beam-focusing gain (Fig. \ref{fig:setup}). Initially, we measured the fabricated lens’s focusing gain $\left(\gamma\right)$ and its beam waist at 3.6 GHz in an anechoic chamber. Then, we tested the lens within the Rydberg atom-based RF receiver setup, comparing EIT splitting with and without the lens at 2.2~GHz and 3.6~GHz.
To characterize the lens inside an anechoic chamber, we used an ETS-Lindgren model 3102 conical log spiral antenna with 3.5~dBi gain at 3.6~GHz, engaging both polarizations simultaneously. The antenna was placed at a 3.2~m distance to ensure that the far-field conditions ar met. Then, a coaxial line near-field probe was used to capture the E-field at various positions on the surface in a symmetrical x and y axes matrix, and up to 50~mm distance away in the z-axis. Results are shown in Fig. \ref{fig:chamber}. We measured $\gamma$ of up to 8.42~dB at the focal point of the lens, gradually decreasing in x,y, and z directions away from it, in line with the theoretical diffraction limit \cite{bEhsan}.  
The atom-based Rydberg RF receiver, described in this work and shown in Fig. \ref{fig:setup}, consists of:
\begin{itemize}
    \item Thorlabs GC19075-CS Quartz vapor reference cell (19 mm diameter x 75 mm length)
    \item Thorlabs PDB250A2 balanced photodiode
    \item 852~nm laser: Toptica DL Pro850 narrow linewidth external-cavity diode laser (ECDL)
    \item 509~nm laser: Toptica DL Pro narrow linewidth ECDL seed at 1018~nm feeding an Azurlight high power fiber amplifier with a second harmonic generation (SHG) doubler head to ~509nm
\end{itemize}
\par In this setup, the RF detection is performed with a balanced photodiode. The probe beam is then split by using a calcite beam displacer, where one beam is then counter-propagated with the coupling laser, and the other one traverses the cell as a reference. Aiming to emulate a typical linearly-polarized E-field source, we separately generated a 2.2~GHz and a 3.6~GHz unmodulated signal with 11~dBm transmit power and the Astro Antenna Ltd. AHA-118S double ridged horn antenna with $\approx$ 2~dBi and 5~dBi gain at these frequencies. The antenna was placed at a 2.2~m distance, and the lens was placed 24~mm away from the vapor cell's center. The antenna was tilted 45$^\circ$ to create a slant angle, allowing the lens to enhance the RF signal in both polarizations. The experimental measurement result is shown in Fig. \ref{fig:eit}. It can be seen that the amount of EIT splitting has effectively doubled with the lens at both 2.2~GHz and 3.6~GHz, therefore increasing the sensitivity of the Rydberg RF receiver over an ultrawide bandwidth of the lens. 

In this letter, we have considered the application of metamaterial GRIN lens technology to enhance the sensitivity of atom-based Rydberg RF receivers without compromising the measurement accuracy by RF parasitics, such as spurious or harmonic emissions. We 3D printed a Luneburg-type GRIN metamaterial lens and demonstrated its effectiveness at increasing the sensitivity of the Rydberg RF receiver over an ultrawide bandwidth of the lens. The proposed solution is low-cost and suitable for various applications, including electric-field metrology, quantum radar, and wireless communications. 

This work was supported in part by the Defence Science and Technology Laboratory (DSTL), which contributed to the construction of the Rydberg RF receiver at PA Consulting, and in part by the NSF-EPSRC grant 2152617.

\section*{\label{sec:level1}Data Availability}
The data that supports the findings of this study
are openly available on GitHub.com in \footnotemark[1].

\footnotetext[1]{
https://github.com/DREMCLTD/Rydberg-RF-sensitivity-experiment.}

\nocite{*}

\section*{\label{sec:level1}References}

\bibliography{aipsamp.bib}

@article{bHarris,
  title = {{Nonlinear optical processes using electromagnetically induced transparency}},
  author = {{Harris, S. E., et al.}},
  journal = {Phys. Rev. Lett.},
  volume = {64},
  issue = {10},
  pages = {1107--1110},
  numpages = {0},
  year = {1990},
  month = {Mar},
  publisher = {American Physical Society},
  doi = {10.1103/PhysRevLett.64.1107},
  url = {https://link.aps.org/doi/10.1103/PhysRevLett.64.1107}
}

@Article{bCai2022,
AUTHOR = {{M. Cai, et al.}},
TITLE = {{Sensitivity Improvement and Determination of Rydberg Atom-Based Microwave Sensor}},
JOURNAL = {Photonics},
VOLUME = {9},
YEAR = {2022},
NUMBER = {4},
ARTICLE-NUMBER = {250},
URL = {https://www.mdpi.com/2304-6732/9/4/250},
ISSN = {2304-6732}
}

@ARTICLE{bHolloway2014,
  author={{C. L. Holloway, et al.}},
  journal={IEEE Transactions on Antennas and Propagation}, 
  title={{Broadband Rydberg Atom-Based Electric-Field Probe for SI-Traceable, Self-Calibrated Measurements}}, 
  year={2014},
  volume={62},
  number={12},
  pages={6169-6182}
}

@article{bKevin2016,
    author = {{C. G. Wade, et al.}},
    title = {{Real-time near-field terahertz imaging with atomic optical fluorescence}},
    journal = {Nature Photon.},
    volume = {11},
    pages = {40-43},
    year = {2017}
}

@article{bKuzmich,
    author = {{L. Li and A. Kuzmich}},
    title = {Quantum memory with strong and controllable Rydberg-level interactions},
    journal = {Nat. Commun.},
    volume = {7},
    number = {13618},
    year = {2016}
}

@article{bSlowlight,
  title = {{Slowing down the speed of light using an electromagnetically-induced-transparency mechanism in a modified reservoir}},
  author = {{R. Liu, et al.}},
  journal = {Phys. Rev. A},
  volume = {96},
  issue = {5},
  pages = {053823},
  numpages = {5},
  year = {2017},
  month = {Nov},
  publisher = {American Physical Society},
  doi = {10.1103/PhysRevA.96.053823},
  url = {https://link.aps.org/doi/10.1103/PhysRevA.96.053823}
}

@article{bGiraldo,
  title = {{State-selective electromagnetically induced transparency for quantum error correction in neutral atom quantum computers}},
  author = {{F. Giraldo, et al.}},
  journal = {Phys. Rev. A},
  volume = {106},
  issue = {3},
  pages = {032425},
  numpages = {14},
  year = {2022},
  month = {Sep},
  publisher = {American Physical Society},
  doi = {10.1103/PhysRevA.106.032425},
  url = {https://link.aps.org/doi/10.1103/PhysRevA.106.032425}
}

@article{bMenchetti,
    author = {{M. Menchetti, et al.}},
    title = {{Digitally encoded RF to optical data transfer using excited Rb without the use of a local oscillator}},
    journal = {Journal of Applied Physics},
    volume = {133},
    number = {1},
    pages = {014401},
    year = {2023}
}

@article{bOtto2021,
    author = {{J. S. Otto, et al.}},
    title = {{Data capacity scaling of a distributed Rydberg atomic receiver array}},
    journal = {Journal of Applied Physics},
    volume = {129},
    number = {15},
    pages = {154503},
    year = {2021}
}

@ARTICLE{bEMC,
  author={{C. L. Holloway, et al.}},
  journal={IEEE Transactions on Electromagnetic Compatibility}, 
  title={{Atom-Based RF Electric Field Metrology: From Self-Calibrated Measurements to Subwavelength and Near-Field Imaging}}, 
  year={2017},
  volume={59},
  number={2},
  pages={717-728},
  doi={10.1109/TEMC.2016.2644616}}

@article{bRadar,
  title = {{Rydberg-atom-based system for benchmarking millimeter-wave automotive radar chips}},
  author = {{S. Bor\'owka, et al.}},
  journal = {Phys. Rev. Appl.},
  volume = {22},
  issue = {3},
  year = {2024},
  month = {Sep},
  publisher = {American Physical Society}
}

@ARTICLE{bBussey,
  author={{L. W. Bussey, et al.}},
  journal={IEEE Sensors Letters}, 
  title={{Quantum Shot Noise Limit in a Rydberg RF Receiver Compared to Thermal Noise Limit in a Conventional Receiver}}, 
  year={2022},
  volume={6},
  number={9},
  pages={1-4},
  doi={10.1109/LSENS.2022.3203465}}

@article{bShot2015,
year = {2015},
month = {sep},
publisher = {IOP Publishing},
volume = {48},
number = {20},
pages = {202001},
author = {{H. Fan, et al.}},
title = {{Atom based RF electric field sensing}},
journal = {Journal of Physics B: Atomic, Molecular and Optical Physics}
}

@ARTICLE{bMaci,
  author={{I. Gashi, et al.}},
  journal={IEEE Transactions on Antennas and Propagation}, 
  title={{GO Analysis of GRIN Lens Antennas by Combining in a Single ODE, Field and Wavefront-Curvature Transport to the Ray Tracing}}, 
  year={2024},
  volume={72},
  number={3},
  pages={2147-2160}
}

@article{bSRR,
    author = {{C. L. Holloway, et al.}},
    title = {{Rydberg atom-based field sensing enhancement using a split-ring resonator}},
    journal = {Applied Physics Letters},
    volume = {120},
    number = {20},
    pages = {204001},
    year = {2022}
}

@article{bSensitivity,
    author = {{N. Prajapati, et al.}},
    title = {{Sensitivity comparison of two-photon vs three-photon Rydberg electrometry}},
    journal = {Journal of Applied Physics},
    volume = {134},
    number = {2},
    pages = {023101},
    year = {2023}
}

@ARTICLE{bSRR2,
  author={{G. Sandidge, et al.}},
  journal={IEEE Transactions on Microwave Theory and Techniques}, 
  title={{Resonant Structures for Sensitivity Enhancement of Rydberg-Atom Microwave Receivers}}, 
  year={2024},
  volume={72},
  number={4},
  pages={2057-2066}
}

@article{bLasers1,
  title = {Enhancing the Sensitivity of Atom-Based Microwave-Field Electrometry Using a Mach-Zehnder Interferometer},
  author = {{W. Yang, et al.}},
  journal = {Phys. Rev. Appl.},
  volume = {19},
  issue = {6},
  pages = {064021},
  numpages = {6},
  year = {2023},
  month = {Jun},
  publisher = {American Physical Society},
  doi = {10.1103/PhysRevApplied.19.064021}
}

@ARTICLE{bLasers2,
  author={{B. Wu, et al.}},
  journal={IEEE Transactions on Antennas and Propagation}, 
  title={Enhancing Sensitivity of Atomic Microwave Receivers Based on Optimal Laser Arrays}, 
  year={2025},
  volume={73},
  number={2},
  pages={793-806}
}

@ARTICLE{bSRR3,
  author={{B. Wu, et al.}},
  journal={IEEE Transactions on Antennas and Propagation}, 
  title={{Enhancing Sensitivity of an Atomic Microwave Receiver via a Fabry-Perot Cavity}}, 
  year={2025},
  volume={73},
  number={2},
  pages={863-872}
}

@article{bEhsan,
    author = {{Hosseininejad, S.E., et al.}},
    title = {{Reprogrammable Graphene-based Metasurface Mirror with Adaptive Focal Point for THz Imaging}},
    journal = {Sci. Rep.},
    volume = {9},
    number = {2868},
    year = {2019}
}

\end{document}